\documentclass[a4paper,showpacs,superscriptaddress,preprint]{revtex4}
\usepackage{hyperref}  
\usepackage{epsfig}
\usepackage{graphics}
\usepackage{color}

\def\bra#1{\langle #1 \vert}
\def\ket#1{\vert #1 \rangle}

\def\imagi{\mbox{\rm i}}
\def\beq{\begin{equation}}
\def\eeq{\end{equation}}

\def\reff#1{(\ref{#1})}

\begin{document}

\title{Harmonic generation by atoms in circularly polarized two-color laser fields with coplanar polarizations and commensurate frequencies }
\author{F. Ceccherini} 
\email{ceccherini@df.unipi.it}  
\affiliation{Istituto Nazionale per la  Fisica della Materia (INFM), sez. A, Dipartimento di Fisica ``Enrico Fermi'', Universit\`a di Pisa, Via F. Buonarroti 2, 56127 Pisa, Italy}
\author{D. Bauer}
\affiliation{Theoretical Quantum Physics (TQP), Darmstadt University of Technology, Hochschulstr.\ 4A, D-64289 Darmstadt, Germany}
\author{F. Cornolti}
\affiliation{Istituto Nazionale per la Fisica della Materia (INFM), sez. A, Dipartimento di Fisica ``Enrico Fermi'', Universit\`a di Pisa, Via F. Buonarroti 2, 56127 Pisa, Italy}

\date{\today}

\begin{abstract} 
The generation of harmonics by atoms or ions in a two-color, coplanar field configuration with commensurate frequencies is investigated through both, an analytical calculation based on the Lewenstein model and the numerical {\em ab initio} solution of the time-dependent Schr\"odinger equation of a two-dimensional model ion.
Through the analytical model, selection rules for the harmonic orders in this field configuration, a generalized cut-off for the harmonic spectra, and an integral expression for the harmonic dipole strength is provided. The numerical results are employed to test the predictions of the analytical model.
The scaling of the cut-off as a function of both, one of the laser intensities and frequency ratio $\eta$, as well as entire spectra for different $\eta$ and laser intensities are presented and analyzed. The theoretical cut-off is found to be an upper limit for the numerical results. Other discrepancies between analytical model and numerical results are clarified by taking into account the probabilities of the absorption processes involved.
\end{abstract}


\maketitle

\section{Introduction}
The possibility of obtaining high frequency radiation through the interaction of a laser field and an atom is a topic that has been extensively addressed during the last two decades from both a theoretical and an experimental point of view (see \cite{joachsal} for recent reviews). Harmonic generation from sources other than atoms, like linear molecules \cite{bandrauk},  ring molecules (e.g. benzene) \cite{alon,vitali}, nanotubes \cite{alonnanotube}, and plasmas \cite{linde} have been also investigated.

From the invariance of the Hamiltonian under dynamical symmetry operations, selection rules for harmonic generation can be elegantly derived \cite{alon,ceccherini}. It turns out that apparently very different target and field configurations  yield selection rules of the same type. Let us consider, e.g.,  a ring molecule with $N$ ions (e.g., $N=6$ in the case of benzene) in a circularly polarized laser field of frequency $\omega$.  The electric field vector lies in the plane that is spanned by the molecule and which we parameterize through the polar coordinates $\rho$ and $\varphi$. The corresponding Hamiltonian (with the laser interaction taken in dipole approximation) is invariant under the dynamical symmetry operation 
\beq   \hat{P}_N = \left(\rho\to\rho, \varphi\to\varphi+\frac{2\pi}{N}, t\to\frac{2\pi}{N\omega}\right) .\label{dso}\eeq
From this invariance follows \cite{alon,ceccherini} that only harmonics of order 
\beq n=gN\pm 1,\qquad g \in {\cal N}_+ \label{selecrule}\eeq 
can be emitted. The harmonic radiation is circularly polarized and subsequent harmonics are alternately clockwise and counter-clockwise polarized. More complicated selection rules arise when also excited states are taken into account \cite{ceccherini}. Let us now turn to the actual target and field configuration examined in the present paper, namely the situation of an atom (or ion) in a circularly polarized two-color laser field of frequencies $\omega$ and $\eta\omega$ with $\eta$ a positive integer number, and coplanar polarizations. In the case of counter-rotating electric field vectors the Hamiltonian is invariant under the same symmetry operation \reff{dso} with $N$ replaced by $\eta+1$. The polar coordinates $\rho$ and $\varphi$ are with respect to the polarization plane now. In the case of co-rotating electric field vectors $N$ has to be replaced by $\eta-1$. 

An appealing feature of the selection rule \reff{selecrule} is the fact that with increasing $N$, i.e., number of ions or frequency ratio of the two laser fields, respectively, less harmonics are emitted within a fixed frequency interval. This filtering effect may be accompanied with more efficient emission of harmonics at short wavelengths which are of interest in spectroscopic applications, for instance.

Harmonic generation in two-color fields has been studied both experimentally \cite{eichmann} and theoretically \cite{becker,milos00,tong}.

The present paper is organized as follows: In Sec. \ref{stuff} the theoretical modeling  proposed by Lewenstein {\em et al.} \cite{lewen0} for harmonic generation in the case of an atom interacting with a single linearly polarized field is extended to the two-color configuration, and several expected features of the harmonic spectra are deduced, among them the scaling of the cut-off and the dependence of the relative dipole strengths within a certain harmonic couple $g$.   In Sec. \ref{numerics} {\em ab initio} numerical results obtained through the integration of the time-dependent Schr\"odinger equation for a two-dimensional model ion are presented and compared with the predictions by the analytical model. Finally, a conclusion is given in Sec. \ref{summary}. 

Atomic units (a.u.) are used throughout the paper. 

\section{Analytical theory} \label{stuff}
A theory of harmonic generation should answer mainly two fundamental questions: (i) which harmonics are emitted and (ii) which is the intensity of the harmonics as a function of the laser and target parameters. These questions have been addressed in \cite{lewen0} for the case of an atom (in the single-electron approximation) interacting with a linearly polarized laser field (in dipole approximation). In a similar approach, the more general and more complicated case of elliptical polarization was studied in \cite{antoine} (single color). The elliptically polarized two-color field was addressed in \cite{milos01} but the discussion of the cut-off law as well as the presentation of the numerical results were restricted to linear polarization there.  
Here, we focus on the case of two laser fields with circular polarizations and arbitrary integer frequency ratios and compare carefully the model predictions with {\em ab initio} numerical simulations.

The electric field caused by the two lasers of frequency $\omega_1=\omega$ and $\omega_2=\eta\omega$ is assumed to be
\beq
\vec{E}(t) = \left(E_x(t), E_y(t), E_z(t) \right ) = \left (E_1\cos(\omega t) + E_2\cos(\eta\omega t), E_1\sin(\omega t) - E_2\sin(\eta\omega t), 0 \right) \label{E}
\eeq
where $E_1$ and $E_2$ are the amplitude of the first and the second laser field, respectively, and the dipole approximation is applied. The two fields are oppositely polarized and coplanar. The case of co-rotating electric field vectors will be discussed later-on.
The vector potential $\vec{A}(t)=-\int^t E(t') \, dt'$ reads
\begin{equation}
 \vec{A}(t) = \left (A_x(t), A_y(t), A_z(t) \right ) = - \left (\frac{E_1}{\omega}\sin(\omega t) + \frac{E_2}{\eta \omega}\sin(\eta \omega t), -\frac{E_1}{\omega}\cos(\omega t) + \frac{E_2}{\eta \omega}\cos(\eta \omega t), 0 \right ). \label{A}
\end{equation}

Our starting point is the dipole moment along the direction $\vec{n}$ as it is calculated in the Lewenstein model (cf.\ Eq.~(10) of Ref.~\cite{lewen0})
\beq x_{\vec{n}}(t)= \imagi \int_0^t dt' \int d^3p\,  \vec{n}\cdot \vec{d}^*(\vec{p}-\vec{A}(t)) \vec{E}(t') \cdot \vec{d}(\vec{p}-\vec{A}(t')) \exp[-\imagi S(\vec{p},t,t')] + \mbox{\rm c.c.} . \label{lewenresult}\eeq
Here, 
\begin{equation}
\vec{d}(\vec{p}) = \bra{\vec{p}}\vec{r}\ket{0},
\end{equation}
and the action $S(\vec{p},t,t')$ is given by
\begin{equation}
S(\vec{p}, t, t') = \int_{t'}^t \left (\frac{(\vec{p} - \vec{A}(t''))^2}{2} + I_p \right )dt''.
\label{action_1}
\end{equation}
In order to arrive at expression \reff{lewenresult} several assumptions have been made in \cite{lewen0}:
({\rm i}) among the bound states only the ground state plays a role in the evolution of the system; ({\rm ii}) the depletion of the ground state can be neglected; ({\rm iii}) in the continuum $V(\vec{r})$ plays no role and the electron is treated like a free particle and can be therefore described py plane waves $\ket{\vec{p}}$; and ({\rm iv}) contributions from continuum-continuum transitions to harmonic generation can be neglected.

For the field \reff{E}, \reff{A}, the general expression \reff{lewenresult} evaluated for, e.g.,  $\vec{n}=\vec{e}_x$, reads
\beq
 x(t) =  {\rm i} \int_0^t  d t' \int d^3 p\, \left (E_x(t')d_x (\vec{p} - \vec{A}(t')  )+ E_y(t')d_y(\vec{p} - \vec{A}(t')\right ) d_x^*(\vec{p}-\vec{A}(t)) \exp[-\imagi S(\vec{p},t,t')]  \eeq 
(the ``$+$ c.c.'' is suppressed from now on). 

The integration over $\vec{p}$ is performed approximately by means of the saddle-point method, assuming   that the major contribution to the integral is given  by stationary points of the classical action, i.e, the points  $p_{x}^{\rm st}$, $p_{y}^{\rm st}$ that satisfy
\begin{equation}
\vec{\nabla}_{\vec{p}} \, S(\vec{p},t,t') = \vec{0}.
\end{equation}
One finds
\begin{equation}
 p_{x}^{\rm st}(t,\tau) = \frac{E_1}{\omega^2 \tau}\bigg( \cos(\omega t) - \cos(\omega (t-\tau)) \bigg ) +  \frac{E_2}{\eta^2 \omega^2 \tau} \bigg ( \cos(\eta \omega t) - \cos(\eta \omega (t - \tau)) \bigg),
\end{equation}
\begin{equation}
 p_{y}^{\rm st}(t,\tau) = \frac{E_1}{\omega^2 \tau}\bigg( \sin(\omega t) - \sin(\omega(t-\tau)) \bigg ) - \frac{E_2}{\eta^2 \omega^2 \tau} \bigg ( \sin(\eta \omega t) - \sin(\eta \omega (t - \tau)) \bigg )
\end{equation}
where $\tau = t - t'$ is the electron's ``travel time.''\ \,  
Introducing the stationary action
\begin{equation}
S_{\rm st}(t, \tau) = S(\vec{p}_{\rm st},t,t-\tau) = \int_{t - \tau}^t \left (\frac{(\vec{p}_{\rm st}(t,\tau) - \vec{A}(t''))^2}{2} + I_p \right )dt''
\label{action_2}
\end{equation}
with $\vec{p}_{\rm st} = (p_x^{\rm st}, p_y^{\rm st},0)$
we obtain,  after the saddle-point integration over $\vec{p}$,
\begin{eqnarray}
 x(t) &=& {\rm i}\int_0^{\infty} d\tau\, \left (\frac{\pi}{\epsilon + {\rm i}\tau/2} \right )^{3/2} \, {\rm exp}(-{\rm i}S_{\rm st}(t, \tau))\, d_x^*(\vec{p}_{\rm st}(t,\tau) - \vec{A}(t))\nonumber \\
&&{\times} \bigg (E_x(t-\tau) d_x(\vec{p}_{\rm st}(t,\tau) - \vec{A}(t-\tau))
+ E_y(t-\tau) d_y(\vec{p}_{\rm st}(t,\tau) - \vec{A}(t-\tau)) \bigg ).
\label{x}
\end{eqnarray}
The factor with infinitesimal $\epsilon$ in \reff{x} comes from the regularized Gaussian integration over $\vec{p}$. It expresses quantum diffusion of the released wave packet and damps away contributions from times $\tau$ much larger than a laser cycle, allowing for the extension of the $\tau$-integration to infinity  \cite{lewen0}.

The stationary action \reff{action_2} can be written in the form
\begin{equation}
S_{\rm st}(t, \tau) = C_0(\tau) + C_1(\tau)\cos \left ( (\eta + 1) \omega \left (t-\tau/2 \right ) \right) 
\label{action_4}
\end{equation}
where $C_0(\tau)$ and $C_1(\tau)$ are given by
\begin{eqnarray}
 C_0(\tau) &=& I_p \tau + \frac{E_1^2}{2\omega^4\tau}(\tau^2 \omega^2 - 2  + 2 \cos(\tau \omega)) + \frac{E_2^2}{2\eta^4 \omega^4 \tau} (\eta^2 \tau^2 \omega^2 -2 + 2 \cos(\eta \tau \omega)),  \\
 C_1(\tau) &=& \frac{4 E_1 E_2}{\eta^2 \omega^4 \tau} \sin(\omega \tau/2) \sin(\eta \omega \tau/2) - \frac{2 E_1 E_2}{\eta (\eta + 1)\omega^3}\sin \left ( (\eta+1) \omega \tau/2 \right).   
\end{eqnarray}
The expression (\ref{action_4}) for the quasi-classical action is very useful and interesting. The time dependence  is given by just one term and through only one effective frequency which is $(\eta + 1)\omega$. This is consistent with the selection rule $g(\eta+1)\pm1$  obtained previously. Taking $\eta = 1$ and $E_1 = E_2 = E/2$ the coefficients $C_0(\tau)$ and $C_1(\tau)$ calculated in \cite{lewen0} for the single linearly polarized field are easily recovered.

As the semi-classical action is the integral over time of the kinetic energy, an expression for the energy gain of the electron is obtained by deriving $S_{\rm st}(t, \tau)$ with respect to $t$, 
\begin{eqnarray}
 \Delta E_{\rm kin}(t, \tau) &=& E_{\rm kin}(t) - E_{\rm kin}(t-\tau) = \frac{\partial S_{\rm st}(t, \tau)}{\partial t} \nonumber \\
  &=& - (\eta + 1) \omega C_1(\tau) \sin \left ( (\eta + 1) \omega \left (\frac{2t-\tau}{2} \right ) \right). 
\end{eqnarray}
The value of the maximum energy gain is equal to the maximum of the function $\tilde{C}_1(\tau)=(\eta + 1) \omega |C_1(\tau)|$. Note, that $C_1(\tau)$ depends on the product of the two electric field amplitudes only. If one of the two fields vanishes the resulting electric field is a pure circularly polarized field and, as expected, there is no possible energy gain and therefore no harmonics are emitted.

We introduce
\begin{equation}
U = \frac{E_1 E_2}{\omega^2},
\label{pondero}
\end{equation}
which, for linear polarization and $E_1=E_2=E/2$, yields the well-known ponderomotive potential $U_p=E^2/4\omega^2$. 
In Fig. \ref{figu1}  the function $\tilde{C}_1(\tau)/U$ is plotted for $\eta = 1, 2, 3, 4$. Writing the maximum energy gain as $\Delta E_{\rm kin}^{\rm max}  = \gamma_{\eta} U$ one obtains for $\eta$ between $1$ and $5$ the values $\gamma_1 = 3.17$, $\gamma_2 = 1.28$, $\gamma_3 = 0.91$,  $\gamma_4 = 0.67$, $\gamma_5 = 0.52$. 

By expanding a part of the integrant in  (\ref{x}) in Fourier components, 
\[
d_x^*(\vec{p}_{\rm st}(t,\tau) - \vec{A}(t)) \bigg (E_x(t-\tau) d_x(\vec{p}_{\rm st}(t,\tau) - \vec{A}(t-\tau)) + E_y(t-\tau) d_y(\vec{p}_{\rm st}(t,\tau) - \vec{A}(t-\tau)) \bigg ) \]
\beq \qquad\qquad\qquad = \sum_M b_M(\tau)\exp(-{\rm i}M \omega t),
\label{bmtau}
\eeq
the dipole projection $x(t)$ becomes 
\begin{eqnarray}
 x(t) &=& {\rm i}\sum_M \int_0^{\infty} d\tau \left (\frac{\pi}{\epsilon + {\rm i}\tau/2} \right )^{3/2} b_M(\tau) \exp(-{\rm i}M \omega t) \nonumber \\ 
&&\times \exp[-{\rm i}C_0(\tau)] \exp \bigg({\rm i}\, C_1(\tau) \cos[(\eta+1)\omega (t -\tau/2)] \bigg).
\label{xt_6}
\end{eqnarray}

In the case of an infinite laser pulse the coefficients $b_M(\tau)$ are non-zero only for $M = -1 + \Delta, 1 + \Delta, -\eta + \Delta$ and $\eta + \Delta$, with $\Delta = m(\eta + 1)$ and $m \in {\cal N}$. This can be easily obtained writing down the dipole $\vec{d}(\vec{p})$ for a hydrogenlike ion \cite{bethe}. Moreover, one can see that $|b_M(\tau)|$ decreases very rapidly with increasing $m$ and the leading terms are those with $m = 0$.  
Making use of a few variable changes and performing a Fourier transformation $x_K=\frac{1}{2\pi}\int_{-\pi/\omega}^{\pi/\omega} dt\, x(t)\exp(\imagi K\omega t)$ yields 

\begin{eqnarray}
 x_K &=& \frac{{\rm i}}{2 \pi}\sum_{M} \int_0^{\infty} d\tau \left (\frac{\pi}{\epsilon + {\rm i}\tau/2} \right )^{3/2} \frac{b_{M}(\tau)}{\eta+1} \exp({\rm i}(K-M) \omega \tau/2) \exp[-{\rm i}C_0(\tau)]\nonumber \\
&& \times \int_{-(\eta+1)\frac{\pi}{\omega}}^{(\eta+1)\frac{\pi}{\omega}} dt \exp \left ({\rm i}\left (\frac{K-M}{\eta +1} \right ) \omega t \right ) \exp \bigg ({\rm i} \, C_1(\tau) \cos(\omega t) \bigg ). 
\end{eqnarray}

The integration over $t$ in zero unless $(K-M)/(\eta + 1)$ is an integer number. In this latter case with the help of the Bessel functions of integer order $J_n(z)$ and taking into account that 

\[
\int_{-\pi}^{\pi} {\rm e}^{{\rm i} z \cos(\theta)} {\rm e}^{{\rm i} n \theta } d \theta = 2 \pi {\rm i}^n J_n(z) 
\]
one obtains 

\begin{eqnarray}
 x_K &=& {\rm i}\sum_{M} \int_0^{\infty} d\tau \left (\frac{\pi}{\epsilon + {\rm i}\tau/2} \right )^{3/2} b_{M}(\tau) \exp[{\rm i}(K-M) \omega \tau/2] \nonumber \\
&& \times \exp[-{\rm i}C_0(\tau)] {\rm i}^{\left (\frac{K-M}{\eta +1} \right )} J_{\frac{K-M}{\eta +1}}[C_1(\tau)].
\end{eqnarray}
Given $(K-M)/(\eta + 1) = g$ with $g \in {\cal N}$ and considering the possible values of $M$ one finds that $x_K$ is different from zero for $K = g(\eta +1) + 1$ and $K = g(\eta +1) - 1$. In the first case the contributing terms are those with $M = 1 + \Delta$ and $M = -\eta + \Delta$ while in the second case they are those with $M = - 1 + \Delta$ and $M = \eta + \Delta$. Hence, 
\begin{eqnarray}
 x_{g(\eta +1) + 1} &=& {\rm i}^{g+1} \int_0^{\infty} d\tau \left (\frac{\pi}{\epsilon + {\rm i}\tau/2} \right )^{3/2} \exp[-{\rm i}C_0(\tau)] \exp[{\rm i}g(\eta +1) \omega \tau/2]  \nonumber \\
&&\times \bigg (b_1(\tau) J_g[C_1(\tau)] + {\rm i} \, \exp[{\rm i}(\eta+1) \omega \tau/2] b_{-\eta}(\tau) J_{g+1}[C_1(\tau)] \bigg ),
\label{xkgp1}\\
 x_{g(\eta +1) - 1} &=& {\rm i}^{g+1} \int_0^{\infty} d\tau \left (\frac{\pi}{\epsilon + {\rm i}\tau/2} \right )^{3/2} \exp[-{\rm i}C_0(\tau)] \exp[{\rm i}g(\eta +1) \omega \tau/2]  \nonumber \\
&&\times \bigg (b_{-1}(\tau) J_g[C_1(\tau)] - {\rm i} \, \exp[{\rm i}(\eta+1) \omega \tau/2] b_{\eta}(\tau) J_{g-1}[C_1(\tau)] \bigg ).
\label{xkgm1}
\end{eqnarray}
By virtue of Eqs.\,\reff{xkgp1} and \reff{xkgm1} it is seen that the selection rule \reff{selecrule} with $N=\eta+1$ is automatically recovered.
Note, that (\ref{xkgp1}) and (\ref{xkgm1}) are made up of two terms, one common (apart for coefficients which are the complex conjugated of each other) proportional to $J_g$, and one proportional to $J_{g+1}$ for (\ref{xkgp1}) and to $J_{g-1}$ for ({\ref{xkgm1}). The coefficients $b_M(\tau)$, to be calculated from (\ref{bmtau}), are functions of the two laser fields $E_1$ and $E_2$. In the limit $E_1 \ll E_2$ one has  $b_{\pm 1}(\tau) \ll b_{\pm \eta}(\tau)$ while $E_1 \gg E_2$  implies $b_{\pm 1}(\tau) \gg b_{\pm \eta}(\tau)$. It follows that the intensities of the two harmonics $g(\eta+1)\pm 1$ in a couple $g$ are expected to be different for $E_1 \ll E_2$ while they converge to the same value for $E_1 \gg E_2$.

\section{\label{numerics} Numerical Simulations}
The numerical simulations were performed by integrating the time-dependent Schr\"odinger equation on a two-dimensional (2D) grid. Reducing the grid to 2D allows to run simulations quickly on every modern PC and does not introduce qualitative modifications to the phenomena we are interested in here. Three-dimensional (3D) simulations are feasible but significantly more demanding. Related work on circular two-color stabilization of H in full 3D has been published recently \cite{bauer}. 

In polar coordinates ($\rho, \varphi$), length gauge, and dipole approximation the time-dependent Schr\"odinger equation under study reads
\begin{eqnarray}
{\rm i}\frac{\partial}{\partial t} \Psi(\rho,\varphi,t) &=& \Bigg [ -\frac{1}{2\rho}\frac{\partial}{\partial \rho} - \frac{1}{2\rho^2}\frac{\partial^2}{\partial \varphi^2} - \frac{\partial^2}{\partial z^2} + V_{\mbox{\scriptsize at}}(\rho) \nonumber \\
&&\quad+ \sin^2(\Theta t) \Big (E_1 \rho \, \mbox{cos}(\varphi -\omega t) + E_2 \rho \, \mbox{cos}\,(\varphi + \eta \omega t) \Big ) \Bigg ] \Psi(\rho,\varphi,t)
\end{eqnarray}
where the two laser pulses have a duration $T=\pi/\Theta$ and a sine-square shape.  $V_{\mbox{\scriptsize at}}(\rho)$ is a ``soft-core'' 2D potential given by\begin{equation}
V_{\mbox{\scriptsize at}}(\rho) = -\frac{\alpha}{\sqrt{\rho^2 + \beta^2}}.
\end{equation}
The parameters $\alpha$ and $\beta$ can be tuned in order to adjust the ionization energy and the ``smoothness'' of the potential. In our simulations we used $\alpha = 2.44$ and $\beta = 0.20$. These values provide an ionization potential of $I_p = 2.0$, i.e., the one of real He${}^+$. The fundamental laser frequency $\omega$ was chosen $0.02 \pi$ and the pulse length was $T=12600$, corresponding to 126 cycles of the frequency $\omega = 1.7$\,eV and $T \approx 300$\,fs. 

Although the details of the model potential do not play a significant role, at least on a qualitative scale, it is useful to know the level scheme in order to understand resonances observed in the numerically obtained harmonic spectra. With the chosen parameters the lowest four excited states have energies $\Omega_1 = 0.985$, $\Omega_2 = 1.375$, $\Omega_3 = 1.548$, and $\Omega_4 = 1.592$.

In Fig.\,\ref{figu2} we present examples of harmonic spectra obtained by Fourier-transforming in time the expectation value $\langle x(t)\rangle = \int\int d\rho \, d\varphi\, \rho \Psi^*(\rho,\varphi,t) \rho\cos\varphi\, \Psi(\rho\varphi,t)$ and plotting the square of the result (hereafter called dipole ``strength''). As expected, the structure of the spectra follows the selection rule, confirming the filtering effect, i.e.,  the number of harmonics present in a certain frequency range decreases with increasing $\eta$. Additional lines of small intensity are also present. Those lines are due either to decays from excited states which are populated during the laser pulse or to recombinations of the electron with states different from the ground state. Such phenomena are also present in the interaction between a laser field and a circular molecule and are extensively discussed in \cite{ceccherini}. These extra lines can be particularly useful for deriving informations about energy shifts because of  the dynamical Stark effect.

In order to verify the analytical findings discussed in the previous section a series of numerical simulations have been performed. In particular, given a fixed value for $E_2$, a series of simulations have been run for different $E_1$ and  $\eta$. In Fig.\,\ref{figu3} the highest resolvable harmonic obtained from the simulations is compared with the highest harmonic expected from the calculations. The interpretation of Fig.\,\ref{figu3} is not straightforward and requires some discussion. Because of angular momentum conservation, in the two-color scheme with opposite polarizations the emission of a harmonic is possible only when $|K_\omega - K_{\eta \omega}| = 1$, where $K_\omega$ and $K_{\eta \omega}$ are the number of photons absorbed from the first and the second laser, respectively. Therefore, in order to achieve an efficient harmonic emission, it is required to find a regime of frequencies and field intensities where the absorption of $K_\omega$ and $K_{\eta \omega}$ photons, respectively, has a reasonably high probability. It follows that the extension of the harmonic spectrum, as predicted by the analytical calculation, plays the role of an upper limit. In fact, the analytical calculation in the previous section does not incorporate the actual absorption processes and their amplitudes, but the electron is rather ``put by hand'' into the continuum. This is the same for the well known case of a single linearly polarized laser where  we have the predicted cut-off at $I_p + 3.17 U_p$. However, this is verified only if the laser frequency and intensity are chosen within proper ranges so that the laser frequency is much smaller than $I_p$, and the ponderomotive energy is comparable or larger than $I_p$. 


Finding the combination of laser parameters that yields the most efficient harmonic generation is not straightforward. However, a first hint about the most promising region in parameter space can be obtained by considering the absorption processes from the two lasers as independent. 
The three plots of the cut-off vs.\ field amplitude $E_1$, shown in Fig.\,\ref{figu3}, were calculated for fixed $E_2=0.16$ and different frequency ratios $\eta=3,4,5$. The agreement between numerical simulations and analytical calculations is particularly good for $\eta = 5$ but less satisfactory for $\eta= 3$ and $\eta = 4$ where the extension of the numerically obtained spectrum  is less than the expected cut-off. By choosing other laser intensities it is possible to have the good agreement for, e.g., $\eta=3$ instead of $\eta=5$. 
However, the set of plots in Fig.\,\ref{figu3} demonstrates that it is possible to obtain harmonic spectra with significant extension even though the laser parameters are not precisely optimized. 
Note, that in Fig.\,\ref{figu3}c the highest observable harmonic in the numerically obtained spectra is slightly greater than the predicted cut-off. This is also well-known from the linearly polarized case and can be attributed to the fact that in the analytical calculations it is assumed that the electron is born and recombines exactly at the origin (where the nucleus is located).  Allowing for offsets from the origin also yields harmonics beyond the calculated cut-offs.

Another feature that should be noted is the asymptotic behavior of the highest observable harmonic order in the numerical simulations. Increasing the electric field amplitude $E_1$, the  probability of absorbing a certain number of photons $K$ from the first laser increases. However, the emission  of high harmonics requires not only the absorption of many photons $K$ from that laser but also the absorption of $K\pm 1$ photons  from the other laser. This second part of the process is the real constraint. In fact, in Fig.\,\ref{figu3} the value of $E_2$ is always constant and the probability of absorbing $K$ photons decreases very rapidly with $K$ and cannot be compensated with the increase of the probability of absorbing $K\pm 1$ photons from the first laser. The same effect can be observed by looking at low values of $E_1$ in Fig.\,\ref{figu3}. Here it is the low probability of absorbing photons of frequency $\omega$ which suppresses the harmonic generation. Summarizing, one can state that the agreement between theory and simulations is good when the absorption processes of order $(I_p + \gamma_\eta U)/(\eta + 1)\pm 1$ have a reasonably high probability for {\em both} laser fields.

The intensity of the emitted harmonics plays, of course, a key role in harmonic generation. Finding a configuration which enhances the efficiency of harmonic generation is important for possible applications because it may allow to use less intense lasers for obtaining a desired radiation intensity. In Fig.\,\ref{figu4} the spectra obtained with three different values of $\eta$ are compared. The harmonics in the $\eta=4$-spectrum are significantly more intense than those obtained with $\eta = 2$ over a wide range of harmonic orders, although the scaling of the theoretical cut-off, i.e., a decreasing cut-off order with increasing $\eta$, may suggest that the opposite should be true. However, our version of the Lewenstein model yields only an upper limit for the cut-off. The fact that higher $\eta$ is favorable here can be understood considering that given a certain harmonic order, the number of photons required for the emission is inverse proportional to $(\eta +1)$, and a lower order absorption process is (for the laser parameters chosen) more likely than a higher one. 

So far, only  harmonic spectra for a given value of the electric field $E_2$ have been discussed. In Fig.\,\ref{figu5} we present three different spectra obtained for different fields $E_2$ and keeping $E_1$ constant. When $E_2$ is low only a few harmonics are present. With increasing field $E_2$ the spectrum assumes a plateau structure. Increasing $E_2$ further leads to violent ionization and, thus, inefficient harmonic generation.  

Another feature predicted by the analytical calculations in the previous section is the intensity of the two harmonics $g(\eta+1)\pm 1$ of a couple $g$ with respect to the two laser intensities. From Eq.(\ref{xkgp1}) and (\ref{xkgm1}) follows that the intensities of the two harmonics should be very different for $E_1 \ll E_2$  and should become very close for $E_1 \gg E_2$. In Fig.\,\ref{figu6} the dipole strengths of the two harmonics in the couple $g=2$ are plotted versus $E_1$ for $\eta=3$ and $\eta=5$. The numerical result clearly confirms the expected behavior. All the harmonics in Fig.\,\ref{figu6} are obtained through the absorption of two photons from the second laser. We have chosen those couples because the total absorbed energy is below the ionization energy and resonance or interference phenomena play a marginal role. 

At low intensities it is possible to consider the absorption from the two lasers as two distinct processes. Therefore we can write the probability $\Gamma (K_\omega, K_{\eta \omega})$ for the overall process as 
\begin{equation}
\label{gamm0}
\Gamma (K_\omega, K_{\eta \omega}) = \Gamma(K_\omega)\Gamma(K_{\eta \omega}). 
\end{equation}
In terms of the generalized cross sections $\sigma_{K_\omega}$ and $\sigma_{K_{\eta \omega}}$ one has
\begin{equation}
\Gamma(K_\omega, K_{\eta \omega}) \propto \sigma_{K_\omega} E_1^{2 K_\omega} \sigma_{K_{\eta \omega}}E_2^{2 K_{\eta \omega}} .
\end{equation}
Note, that in contrast to the single, linearly polarized laser field where the $n$th harmonic perturbatively scales with the $n$th power of the laser intensity $I$, here, in the two-color case, this is not true. Given for example the [$g(\eta + 1) + 1$]th harmonic, it scales with the $(g+1)$th power of the intensity $I_1$ and the $g$th power of the intensity $I_2$.
Consequently, by plotting double-logarithmically the intensity of a certain harmonic versus $E_1$ or $E_2$ one expects a straight line, the slope of which yields information about the number of photons absorbed from that laser. Instead, if no straight line is obtained, the factorization of Eq.\,(\ref{gamm0}) is not valid. In Fig.\,\ref{figu7} the low-intensity region of Fig.\,\ref{figu6}b is plotted on a log-log scale. We see that all the points are indeed aligned along straight lines so that (\ref{gamm0}) is an acceptable approximation there. However, moving towards higher field amplitudes $E_1$ in Fig.\,\ref{figu6} it is obvious that the approximation (\ref{gamm0}) will soon break down.

Finally, let us briefly  discuss the case where the two laser fields have the same circular polarizations and the selection rule $g(\eta -1) \pm 1$ holds \cite{ceccherini}. In order to achieve angular momentum conservation in the emission process of a certain harmonic, the absorption of photons from one field has to be accompanied by the emission of photons of frequency equal to that one of the other laser field \cite{milos00}. According to the selection rule, a frequency ratio $\eta = 5$ in the co-rotating configuration provides a spectrum with the same harmonic orders present as with $\eta = 3$ for counter-rotating electric field vectors.  The two spectra are shown in Fig. \ref{figu8}. We see that in the region around harmonic order $n=50$ (off all resonances) the intensity of the harmonics in the co-rotating case is significantly lower (between one and two orders of magnitude). Given a certain harmonic, the relative intensities can provide useful information for further investigations about the absorption and emission processes.

\section{\label{summary} Conclusion}
We have investigated the generation of harmonics by atoms or ions in the two-color, coplanar field configuration for different values of integer frequency ratio and different laser intensities. Through an analytical calculation based on the Lewenstein model, the selection rule for the harmonic orders in this field configuration, a generalized cut-off for the harmonic spectra, and an integral expression for the harmonic dipole strength has been calculated.

Numerical {\em ab initio} simulations of a two-dimensional model ion subject to the two-color, coplanar field configuration were performed. The numerical results did not suffer from the various assumptions made in the Lewenstein model and therefore served as an important benchmark for the theoretical predictions.

The scaling of the cut-off as a function of both, one of the laser intensities and frequency ratio $\eta$, as well as entire spectra for different $\eta$ and laser intensities were presented and analyzed. The theoretical cut-off was found to be an upper limit for the numerical results. The theoretically predicted relative strength of the two harmonics $g(\eta+1)\pm1$ in a certain couple $g=1,2,3,\ldots$ when one laser is much more intense than the other was confirmed by the numerical simulations. The dipole strength of the harmonics in general increase with increasing $\eta$ although the scaling of the theoretical cut-off, i.e., a decreasing cut-off order with increasing $\eta$, may suggest the opposite. This was found to be due to the decreasing order of the absorption processes involved.

Further studies can be undertaken in order to investigate the importance of resonances within this scheme and the possibility of using such resonances for enhancing strongly the efficiency of certain harmonic lines.       

\begin{acknowledgments}
This work was supported in part by INFM through the Advanced Research Project CLUSTERS. Useful discussions with A. Macchi and  N. Davini, as well as the possibility of using the calculation facility at PC\,${}^2$ in Paderborn, Germany, are  gratefully acknowledged.
\end{acknowledgments}

\pagebreak

\begin{figure}
\begin{center}
\epsfbox{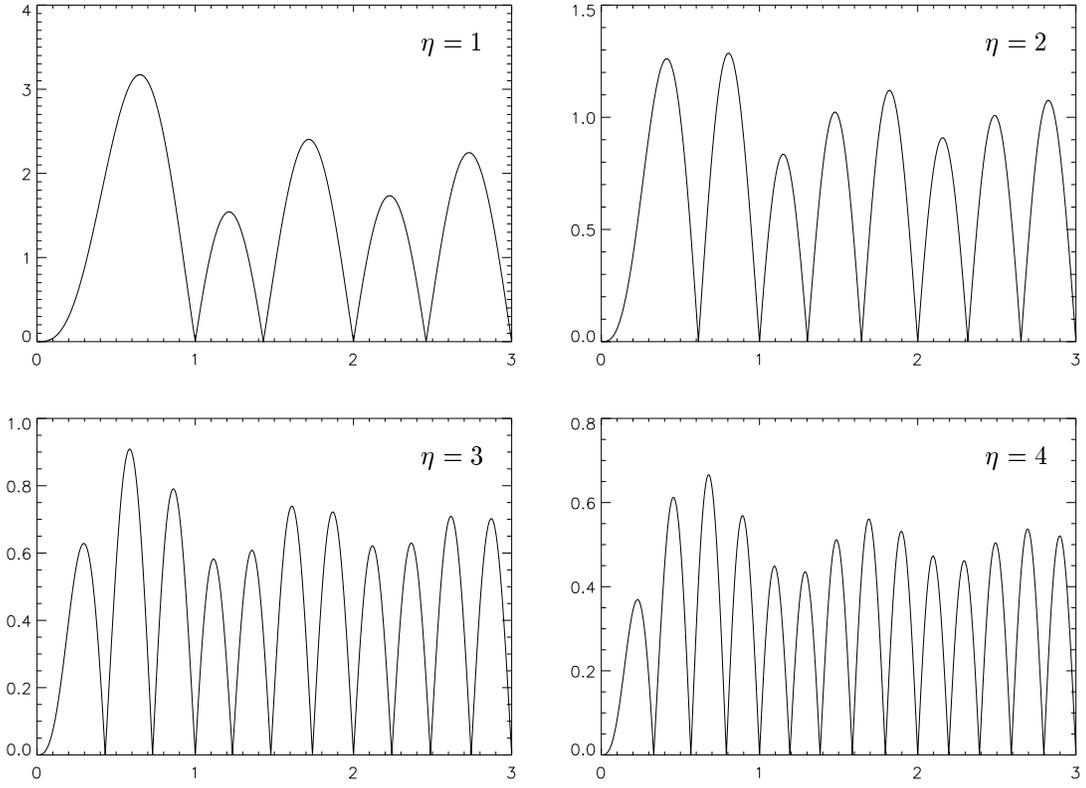}
\end{center}
\caption{\label{figu1} The function $\tilde{C}_1(\tau)/U$ is plotted for $\eta = 1,2,3,4$. The time $\tau$ is counted in periods $2\pi/\omega$. For $\eta = 1 $ the maximum energy gain is, as expected, $3.17\,U_p$, for higher $\eta$ the maximum energy decreases. }
\end{figure}

\pagebreak

\begin{figure}
\begin{center}
\epsfbox{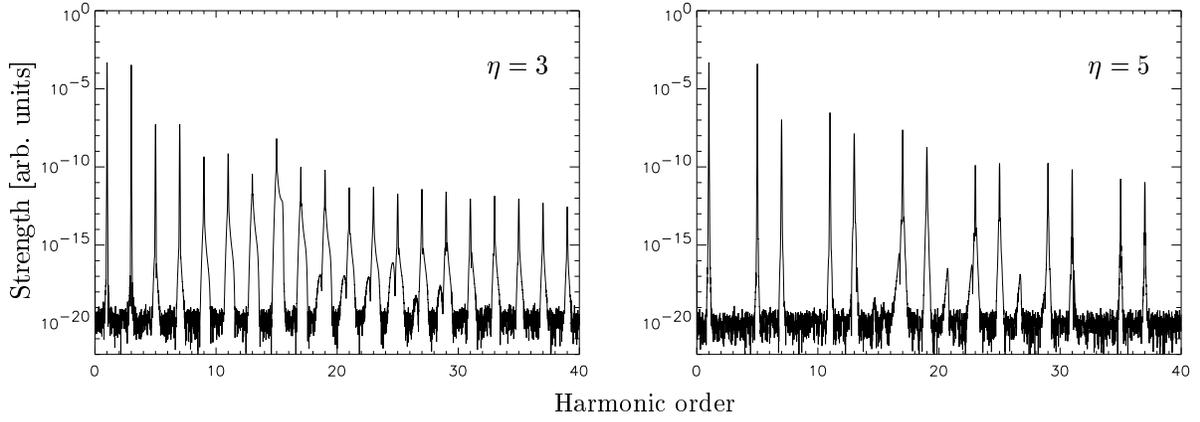}
\end{center}
\caption{\label{figu2} Harmonic spectra for $\eta = 3$ and $\eta=5$. Both spectra have the structure expected from the selection rules. Additional lines between the 15th and the 30th harmonic are due to the population of excited states. The electric fields are $E_1 = 0.16$ and $E_2 = 0.13$. }
\end{figure}

\pagebreak

\begin{figure}
\begin{center}
\epsfbox{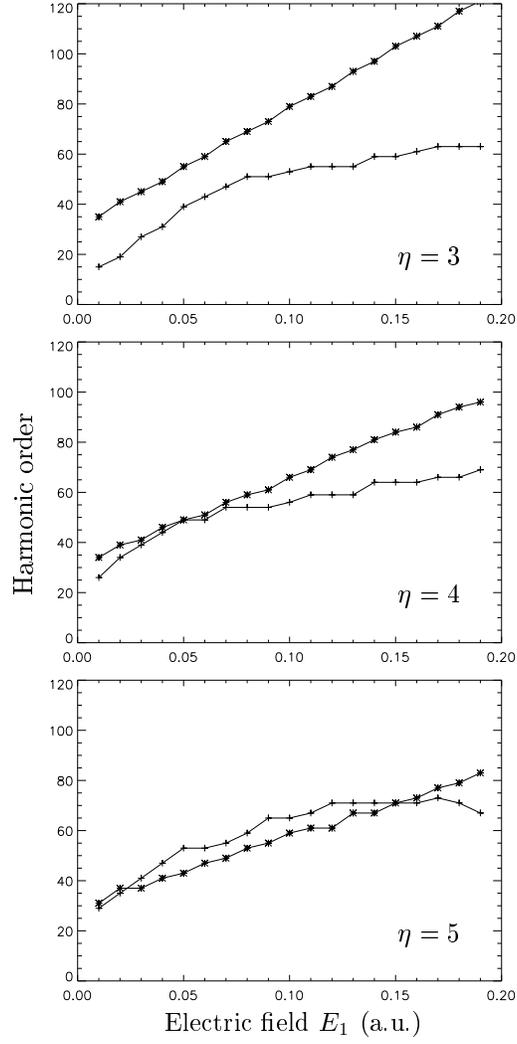}
\end{center}
\caption{\label{figu3} Highest observable harmonic as expected from the Lewenstein-type model ($*$) and from the numerical simulations ($+$) for different values of $\eta$ versus the electric field $E_1$. The electric field $E_2$ is constant, $E_2 = 0.16$. With this particular laser intensity the agreement between the two curves is  good for $\eta = 5$. }
\end{figure}

\pagebreak

\begin{figure}
\begin{center}
\epsfbox{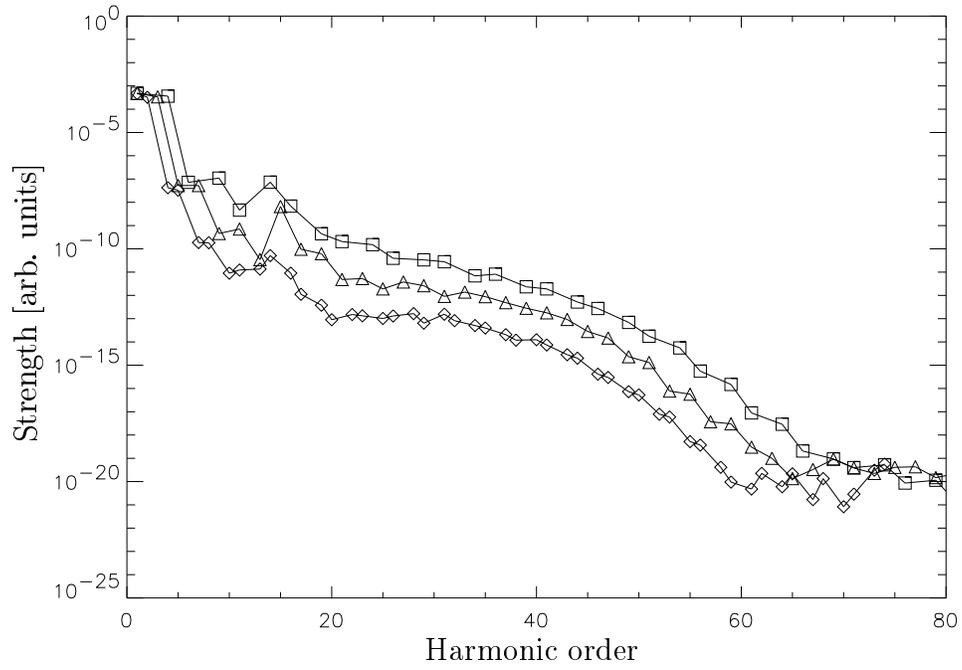}
\end{center}
\caption{\label{figu4} Harmonic spectra for different values of $\eta$. Diamonds: $\eta = 2$, triangles: $\eta = 3$, squares: $\eta = 4$. The laser fields were $E_1 = 0.13$ and  $E_2 = 0.16$. In all three cases the harmonic spectra show a similar structure. The conversion efficiency increases with $\eta$. }
\end{figure}

\pagebreak

\begin{figure}
\begin{center}
\epsfbox{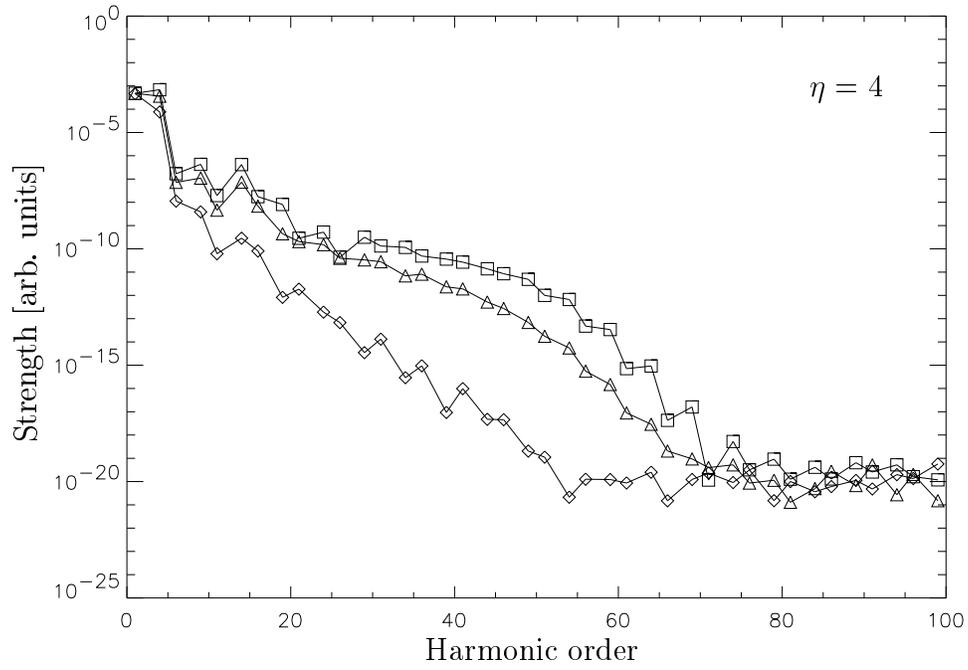}
\end{center}
\caption{\label{figu5} Harmonic spectra for different values of the electric field $E_2$. Diamonds: $E_2 = 0.06$, triangles: $E_2 = 0.13$, squares: $E_2= 0.18$. Th electric field $E_1$ is constant, $E_1 = 0.13$. With increasing value of the electric field $E_2$ a ``plateau-like'' structure in the spectrum appears. }
\end{figure}

\newpage

\begin{figure}
\begin{center}
\epsfbox{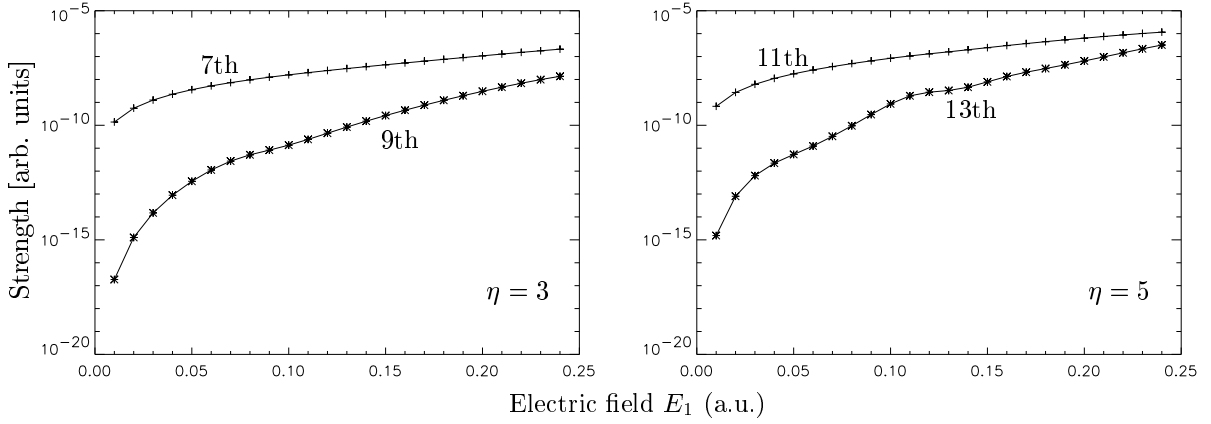}
\end{center}
\caption{\label{figu6} Behavior of the two harmonics belonging to the second couple vs.\ the laser intensity $E_1$. $E_2 = 0.13$ was held constant. While the 7th harmonic for $\eta = 3$ and the 11th harmonic for $\eta = 5$ are given by the absorption of two photons from the second laser and one from the first laser, the 9th harmonic for  $\eta = 3$ and the 13th harmonic for  $\eta = 5$ are generated by the absorption of two photons from the second laser and three from the first one. As expected from the Lewenstein model, with increasing intensity of the electric field $E_1$, the strengths of the two lines of each couple become closer.} 
\end{figure}

\newpage 

\begin{figure}
\begin{center}
\epsfbox{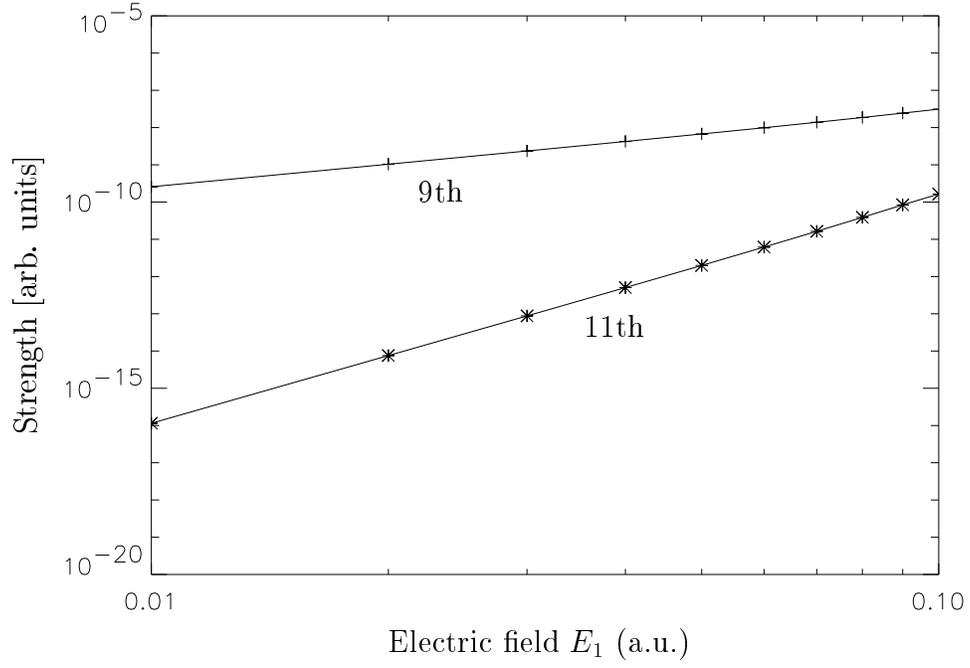}
\end{center}
\caption{\label{figu7} Dipole strength of the two harmonics no.\ 9 and 11 of the second couple ($\eta = 4$) vs.\ the electric field $E_1$. The amplitude $E_2=0.13$ was held constant. All the points are aligned along a straight line on the log-log scale. The slopes are 2 and 6, corresponding to the absorption of one and three photons, respectively.} 
\end{figure}

\pagebreak

\begin{figure}
\begin{center}
\epsfbox{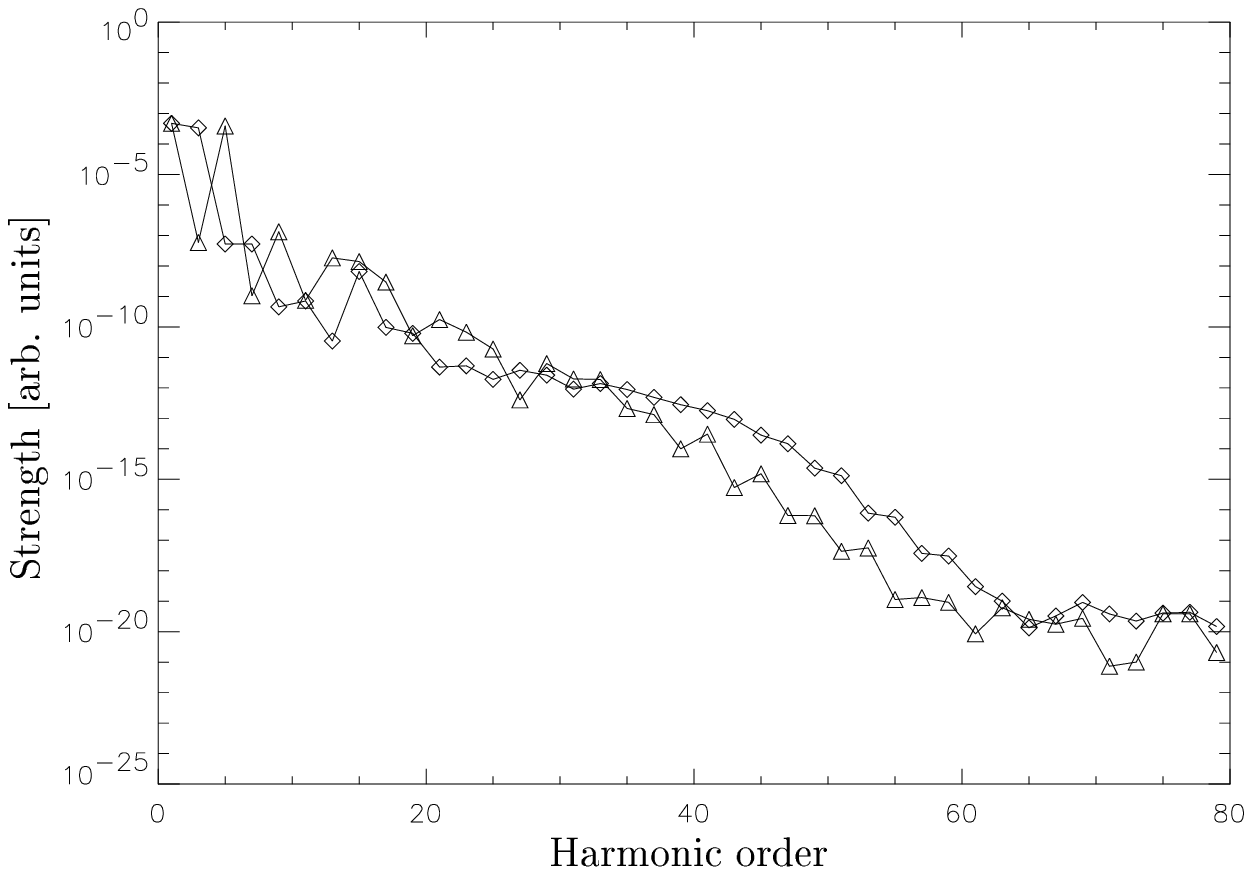}
\end{center}
\caption{\label{figu8} Harmonic spectra for opposite and same polarization of the two laser fields with $\eta$ adjusted in such a way that the same selection rule $4g\pm1$ holds. With $\eta = 3$ (diamonds) the two laser fields have opposite polarization, while for $\eta = 5$ (triangles) the polarization is the same. While the emitted harmonics are the same, the intensity (apart effects due to resonances) is significantly higher in the case where the laser fields have opposite polarization. The electric fields are $E_1=0.16$ and  $E_2=0.13$.}
\end{figure}


\begin{thebibliography}{99}
\bibitem{joachsal} P. Sali\`eres, A. l'Huillier, P. Antoine and M. Lewenstein, Adv.\ At., Mol., Opt., Phys. {\bf 41}, 83 (1999); C. J. Joachain, M. D\"orr, and N. J. Kylstra, Adv.\ At.\ Mol.\ Opt.\ Phys.\ {\bf 42}, 225 (2000); T. Brabec, and F. Krausz, Rev. Mod. Phys. {\bf 72}, 545 (2000).
\bibitem{bandrauk} A. D. Bandrauk and N. H. Shon, Phys. Rev. A, {\bf 66}, 031401(R) (2002).
\bibitem{alon} O. Alon, V. Averbukh and  N. Moiseyev, Phys. Rev. Lett. {\bf 80}, 3743 (1998).
\bibitem{vitali} V. Averbukh, O. Alon, and N. Moiseyev, Phys. Rev. A {\bf 64}, 033411 (2001).
\bibitem{alonnanotube} Ofir E. Alon, Vitali Averbukh, and Nimrod Moiseyev, Phys. Rev. Lett. {\bf 85}, 5218 (2000). 
\bibitem{linde} D. von der Linde and K. Rz\`azewski, Appl. Phys. B {\bf 63}, 499 (1996).
\bibitem{ceccherini} F. Ceccherini, D. Bauer, and F. Cornolti, J. Phys. B {\bf 34}, 5017 (2001).
\bibitem{eichmann} H. Eichmann, A. Egbert, S. Nolte, C. Momma, B. Wellegehausen, W. Becker, S. Long and J. K. McIver, Phys. Rev. A, {\bf 51} R3414 (1995).
\bibitem{becker} S. Long, W. Becker, and R. Kopold, Phys. Rev. A {\bf 52}, 2262 {1995}. 
\bibitem{milos00} D. B. Milosevic, W. Becker and R. Kopold, Phys. Rev. A {\bf 61}, 063403 (2000).
\bibitem{tong} X. M. Tong and S. I. Chu, Phys. Rev. A {\bf 58}, R2656, (1998). 
\bibitem{lewen0} M. Lewenstein, Ph. Balcou, M. Yu. Ivanov, Anne L'Huillier, and P. B. Corkum, Phys. Rev. A {\bf 49}, 2117 (1994). 
\bibitem{antoine} Philippe Antoine, Anne L'Huillier, Maciej Lewenstein, Pascal Sali\`eres, and Bertrand Carr\'e, Phys. Rev. A {\bf 53}, 1725 (1996).
\bibitem{milos01} D. B. Milosevic B. Piraux, Phys. Rev. A {\bf 54}, 1522 (1996).
\bibitem{bethe} H. A. Bethe and E. E. Salpeter, {\em Quantum Mechanics of One and Two Electron Atoms} (Academic, New York, 1957). 
\bibitem{bauer} D. Bauer and F. Ceccherini,  Phys. Rev. A {\bf 66}, 053411 (2002).


\end{thebibliography}
\end{document}